\pdfoutput=1

\def\year{2022}\relax
\documentclass[letterpaper]{article} 
\usepackage{aaai22}  
\usepackage{times}  
\usepackage{helvet}  
\usepackage{courier}  
\usepackage[hyphens]{url}  
\usepackage{graphicx} 
\usepackage{natbib}  
\usepackage{caption} 
\DeclareCaptionStyle{ruled}{labelfont=normalfont,labelsep=colon,strut=off} 
\usepackage{algorithm}
\usepackage{algorithmic}

\usepackage{newfloat}
\usepackage{listings}
\floatstyle{ruled}
\newfloat{listing}{tb}{lst}{}
\floatname{listing}{Listing}

\newcommand*\emoji[1]{\includegraphics[height=8pt]{emojis/#1.png}}

\title{Empathetic AI for Empowering Resilience in Games
}

\author {
    Reza Habibi, \textsuperscript{\rm 1}
    Johannes Pfau, \textsuperscript{\rm 1}
    Jonattan Holmes,\textsuperscript{\rm 1}
    Magy Seif El-Nasr \textsuperscript{\rm 1}
}
\affiliations {
    \textsuperscript{\rm 1} University of California, 1156 High Street, Santa Cruz, CA 95064, USA
\\
    rehabibi, jopfau, joholmes, mseifeln @ucsc.edu
}

\usepackage{bibentry}

\begin{document}

\maketitle

\begin{abstract}
Failure and resilience are important aspects of gameplay. This is especially important for serious and competitive games, where players need to adapt and cope with failure frequently. In such situations, emotion regulation - the active process of modulating ones' emotions to cope and adapt to challenging situations - becomes essential. It is one of the prominent aspects of human intelligence and promotes mental health and well-being. While there has been work on developing artificial emotional regulation assistants to help users cope with emotion regulation in the field of Intelligent Tutoring systems, little is done to incorporate such systems or ideas into (serious) video games. In this paper, we introduce a data-driven 6-phase approach to establish \textit{empathetic artificial intelligence} (\textit{EAI}), which operates on raw chat log data to detect key affective states, identify common sequences and emotion regulation strategies and generalizes these to make them applicable for intervention systems.
\end{abstract}

\section{Introduction}
Remember the last time your team failed after extended periods of struggling on a difficult video game challenge? You and other members of your team may have exhibited distinct emotional responses. You may have started to analyze your strategy or just gave up. Based on psychological theories \cite{sheppes2011emotion}, when facing a challenge or failure, one would often exhibit situation assessment, re-assessment and/or re-planning to respond to events of failure. However, an important part of this adaptation process is emotion regulation \cite{martin2013adaptability}. Emotion regulation is the process by which individuals attempt to manage their emotions in order to regulate their affective and behavioral responses to emotion-eliciting events \cite{mcrae2020emotion,adrian2011methodological,gratz2004multidimensional}. Psychology as well as learning science theories have stressed the importance of this process and demonstrated that effective emotion regulation has a significant impact on higher cognitive flexibility \cite{isen1987positive} and abilities to cope within both individual as well as group activities \cite{cohn2009happiness}. The question we aim to tackle in this paper is how to develop game AI assistant tools that aid players in this emotion regulation phase, thus allowing us to enhance adaptation processes within players and teams. This is especially important for (collaborative) serious games as well as competitive games such as esports. We also hypothesize this may lead to broader impacts on players' resilience and coping abilities outside of games.

Integrating tools to assist users in the emotion regulation phase towards better learning has been a topic of interest in Intelligent Tutoring Systems (ITS), whereby AI agents are developed with the ability to recognize negative feelings such as confusion, frustration, etc., and respond with appropriate emotion regulation strategies \cite{woolf2009affect,d2013autotutor,klein2002computer}. However, none of these works focused on games or team-based activities, where emotion regulation happens at the social level rather than only at the individual level. An example is \cite{woolf2009affect}, where they developed a system called ``Wayang'' that identifies five independent emotional variables, such as frustration, motivation, self-confidence, boredom, and fatigue. It also used text or mirrored student actions in order to regulate their emotions using emphatic responses, displayed changes in the agent's voice and gesture, presented graphs and hints, or gave encouragement. Further, Tian et al. proposed an interactive text-oriented emotion regulation system employing active listening and focusing on providing non-judgmental feedback to emotionally distressed students \cite{tian2014recognizing}. They used text classification to identify and regulate e-learner's emotions based on their textual interactions. Their framework identified emotional patterns after students submitted a message, which is one of the steps in active listening. They also utilized a case-based reasoning algorithm to suggest a similar emotion regulation such as a text-based advice when the student experiences a negative state such as boredom.

In this paper, we propose the utilization of \textit{Empathetic Artificial Intelligence} (\textit{EAI}) into team-based games, developing AI tools to detect and assist players' emotion regulation. We present a work in progress system. Previous work around ITS for emotion regulation used predominantly symbolic rule-based systems adopting psychological theories of emotion regulation ( notably Gross \cite{gross1999emotion}), while we approach the problem using a data-driven technique. We utilize recorded data through user interaction to foster a deeper understanding of the players' emotional and problem-solving processes. In this particular proof of concept, we collect conversation log data containing team emotion regulation behaviors from a team-based game. Using human interpretation, we decipher and label patterns by which players in a team can help each other cope and thus regulate each others' emotions. Our hope is that the patterns can then be used as input to an agent-based AI-assisted intervention system for emotion regulation.

For collecting the data, we use an Alternate Reality Game that we developed called \textit{LUX} \cite{habibi2022data}. \textit{LUX} is a multiplayer team-based cooperative game in which we embed players in a fictional narrative that unfolds through interaction with the real world. In \textit{LUX}, we focused on creating challenges and stressors to evoke failure and emotion regulation. Players interact using chat, as common in team-based games. We collected and de-identified this chat data. This is then further used for our method presented in this paper. It should be noted that chat data is common in other team-based games as well as social media activities. Thus, the introduced approach arguably generalizes onto a wider scope of (team-based) games where communication plays a central role and can be adapted to different types of assistive AIs for emotional regulation beyond \textit{LUX}.

Through the method presented here, we address the following research questions: (1) how can we identify emotion regulation strategies and patterns from chat data and (2) how do these patterns help us develop an empathetic AI (EAI) assistant bot for emotion regulation? In this paper we focus on the first question, leaving question 2 for future work. It is then divided into the following sections: We first review related work, then demonstrate our proposed approach, before discussing limitations and future work.

\section{Related Work}


There is extensive research in the area of Intelligent Tutoring Systems (ITS) on Emotion-sensitive Intelligent Tutoring Systems (EITS), which are approaches that particularly target emotion regulation. In their overarching review, Malekzadeh, Mustafa and Lahsasna focused on emotion regulation and ITS \cite{malekzadeh2015review}, following that if ITS can adapt to the affective state (emotional state) of the learners, it will function as an intervention system and significantly improve its performance. We will review this work in the following. However, to the best of our knowledge, no work has been done on data-driven detection of emotion regulation (around failures), nor has such work been implemented in the context of a game or team-based game activity. 

Prior work on emotion regulation strategies in ITS can be categorized as problem-focused, emotion-focused, and active listening strategies, where problem-focused coping strategies are defined as solving the problem that led to the emotional situation \cite{malekzadeh2015review} and active listening is defined as an effective method for managing emotions by identifying the emotion after receiving an input, such as a sentence. Chaffer et al. used problem-focused strategies within their virtual tutor developed to teach data structure web course \cite{chaffar2009inducing}. In their study, they outlined two different phases: First, the tutor used different course-based examples and definitions to change the situation that caused the negative emotion (problem-focused approach). Secondly, they followed an emotion-focused approach. After providing evaluation results to students, the tutor used \textit{encouragement}, \textit{recommendation}, and \textit{congratulation} as a way to encourage students to improve their performance in the future. Similarly, Zakharov et al. utilized an ITS agent to respond to students' behaviors by defining a set of roles associated with their mental and emotional states. These rules' main goal is to determine the agent's verbal responses as well as emotional appearances. For instance, when they obtain an incorrect response from students, they will provide them with a list of errors accompanied by the proper facial expression. The idea behind their methods is to distract students from negative emotions by showing them their results and move them towards the goal. They conducted the study in a database course with a small group of students and discovered that the presence of an emotion-aware agent beared significant advantages over its non-emotional counterpart.

Tian et al. designed a recommmendation system for emotion regulation that implemented active listening \cite{tian2014recognizing}. Their approach would detect emotions of users through text and predict their emotions using simple classification methods, such as Support Vector Machines or Naive Bayesian Classifiers on courses' chatroom, online Q\&A, and group discussion chat data. The proposed framework analyses the learner's emotion trend by detecting interaction features from user input text. Moreover, it is capable of determining the best emotion regulation strategies. They utilized a database of emotion regulation strategies to offer strategies to students. This database was created by implementing successful emotion regulation strategies based on Gross' emotion regulation strategies \cite{gross2001emotion}. D'mello et al. proposed the \textit{Affective AutoTutor} that assists students in mastering difficult physics concepts by interacting with them in natural language and adaptive dialog similar to human tutoring \cite{d2013autotutor}. Their technology can detect negative emotional states such as irritation and boredom. \textit{Affective AutoTutor} helps students by holding a conversation in natural language, simulating motivational strategies similar to a human tutor and responding to the students' cognitive states. The \textit{affect-sensitive} version of the proposed system which is called \textit{Shakeup} is capable of detecting and regulating negative emotional states, and synthesize emotions of the animated agent. The result of their experiment demonstrated that the \textit{Supportive} version was more effective than the regular one for the students with lower-domain knowledge compared to the more knowledgeable students. Further, Mao and Li \cite{mao2010agent} proposed an emotion agent tutor called \textit{Alice}. \textit{Alice} can recognize emotional states using facial expressions, speech, and text. It could adapt to the learner's emotional state based on facial expression and synthesize speech and text using an Artificial Intelligence Markup Language (AIML) retrieval mechanism. They assisted educators in identifying appropriate educational and emotional responses for each unique scenario applicable to \textit{Alice}. Moreover, they indicated that emotional-aware agents in ITS can enhance the satisfaction of the learning setting for the student.

\section{Approach}

\subsection{\textit{LUX}}
\textit{LUX} is constructed to be a multiplayer team-based cooperative game targeting coping, adaptability and team dynamics within social settings \cite{habibi2022data}. The narrative of the game revolves around the founding principles of the University of California at Santa Cruz (UCSC). UCSC was founded based on the principles of making a Utopian campus in the woods, where every student could thrive in harmony, learn at their own pace without pressure and worry about grades. Using this as a context, the game begins with a hiring message from a group of people who believe in the university's founders' grand vision. They seek support for the formation of a group of loyalists who will restore the university's core values. Upon acceptance of the invitation, players are added to a \textit{Discord} server where they form teams of three members and games unfold in multiple episodes. In each episode, players receive a series of puzzles from a designated \textit{Discord} bot. Each puzzle reveals information about the university's history. In order to solve these puzzles, players are required to find clues distributed around the UCSC main campus. Using an augmented reality (AR) mobile app, players can scan AR markers belonging to the clue, revealing information and leading to the solution of a puzzle.

Data from \textit{LUX} is composed of chat logs collected through \textit{Discord}. We de-identified this data based on our approved IRB protocol. These logs were then used as input for the data-driven methodology discussed next.

 \begin{figure}[htp]
    \centering
    \includegraphics[width=\linewidth]{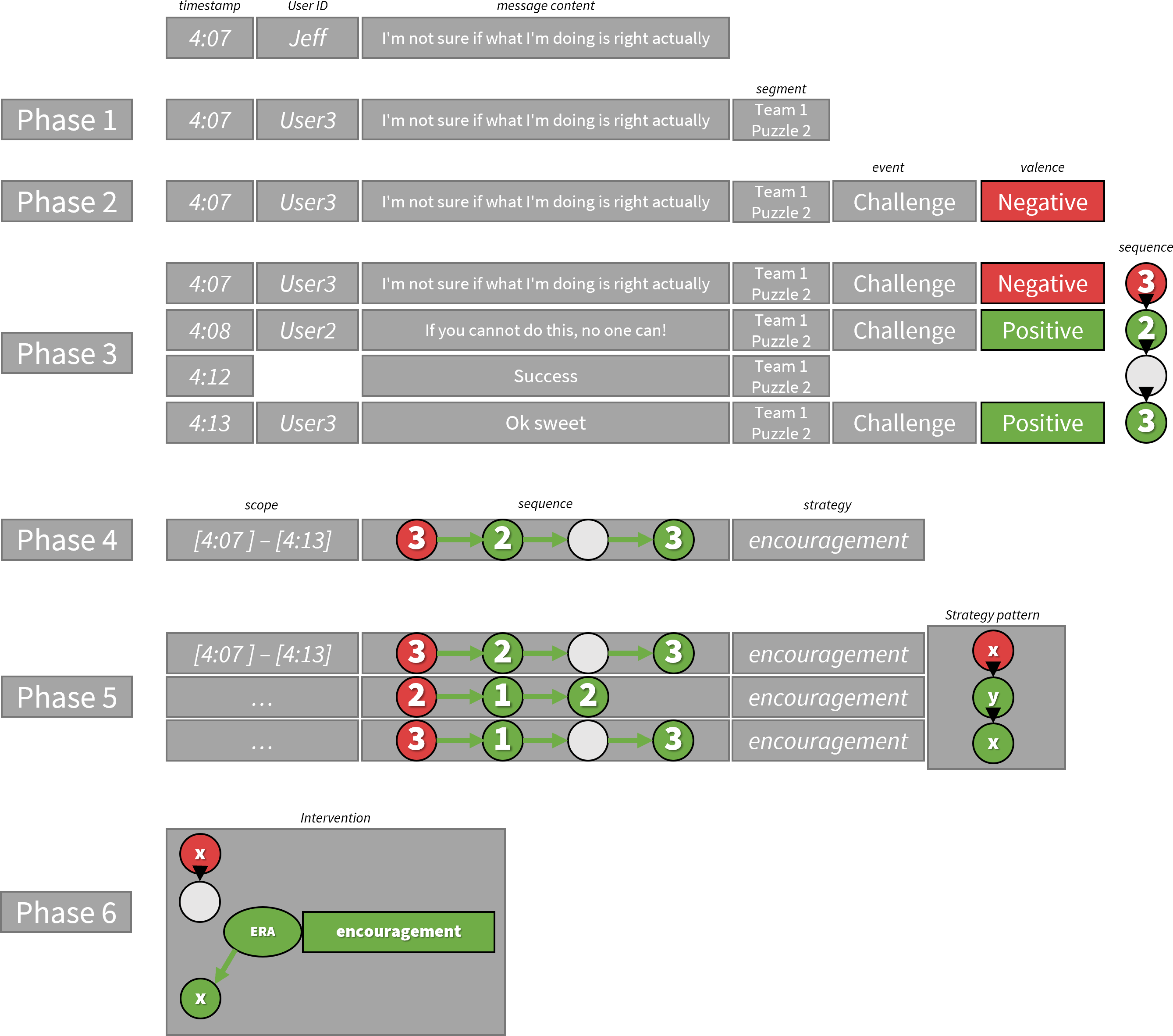}
    \caption{Six-Phase model of extracting and utilizing factors and strategies of emotion regulation}
    \label{fig:SixPhases}
\end{figure}

\subsection{Proposed Data-driven Approach to EAI}
Given data collected from team chat (in this case \textit{LUX}), we use a multi-stage process to identify emotion regulation strategies. Figure \ref{fig:SixPhases} shows the how raw data is processed, abstracted and semanticized. Phase 1 is focused on pre-processing this data to prepare it for the subsequent procedure. Phase 2 then applies semantic labeling and classification into emotional valence, using a set of theory-driven affective categories that can be used across applications. Phase 3 is focused on graphing sequences of these data labels based on frequency and association. Phase 4 uses these directed graph models to perform pattern identification of particular emotion regulation instances. Phase 5 then involves developing generalizable emotion regulation strategies and lastly, phase 6 involves developing situated intervention. Within the work in progress discussed in this paper, we implemented the first four phases, while using a human in the loop approach. Furthermore, we will discuss automation processes of these phases and implementation of phases 5 and 6 as future work.

\subsection{Phase 1: Data Collection and Preparation}
To target semantic labeling of chat data, we first need usable, clean and definite segments of conversation. Arguably, the most usable dialog for detecting emotion regulation stems from interdependent communication between users, which is what we tailor the following approach to. We propose to divide chat logs into episodes based on their topical and temporal affinity. Further, to enhance subsequent sentiment analysis, emojis or shared images are tokenized into textual representations. Eventually, particular lines are pseudonymized to retain user context without identifiable information.


\subsection{Phase 2: Sentiment Analysis \& Labeling}


Given the logs, we labeled the segments of chat data. To prepare the stage for identifying emotion regulation strategies within the chat data in later phases, we developed a simple labeling scheme. In this paper, we focused on the valence of an emotion that constitutes the positive or negative direction of an emotion. 
For the sake of brevity, we neglected the dimension of arousal (i.e., the magnitude or intensity of an emotion), which will be extended in future work. This frugal classification of emotions helped us show how players' feelings changed after an uncertain condition in the game. We also needed contextual details of this emotion expression, since emotion regulation strategies can produce very different outcomes in different contexts and they emerge as a result of uncertain events \cite{gross2001emotion}. For this, we captured the major events of “Failure”, “Challenge”, and “Conflict” based on our game context. Moreover, we registered primary puzzle-based events, such as ``Getting Puzzle'' and ``Success'', while we leave deeper breakdowns for future work. 
Together with ``Positive" and ``Negative" sentiment, these resulted in the labels shown in Table \ref{tab:AffectivityFacets}. Future work will explore the extension of this labeling scheme to incorporate other taxonomies of affect, such as \cite{scott2012adapting} and also situations and strategies discussed by \cite{gross2015emotion}. To have a clear spatial view of emotion regulation, we used a combination of labels such as $<$UserID$><$Pain Point$><$Emotion$>$ for each player.

Given the proposed scheme, the labeling process proceeded according to qualitative research labeling processes, including Inter-Rater Reliability (IRR) calculations to check for reliability of the codes. Eventually, we ensured to use labels with a high IRR (Fleiss’ $\kappa=0.88$) among two independently coding members of the research team \cite{fleiss1971measuring}.

\begin{figure}[htp]
    \centering
    \includegraphics[width=\linewidth]{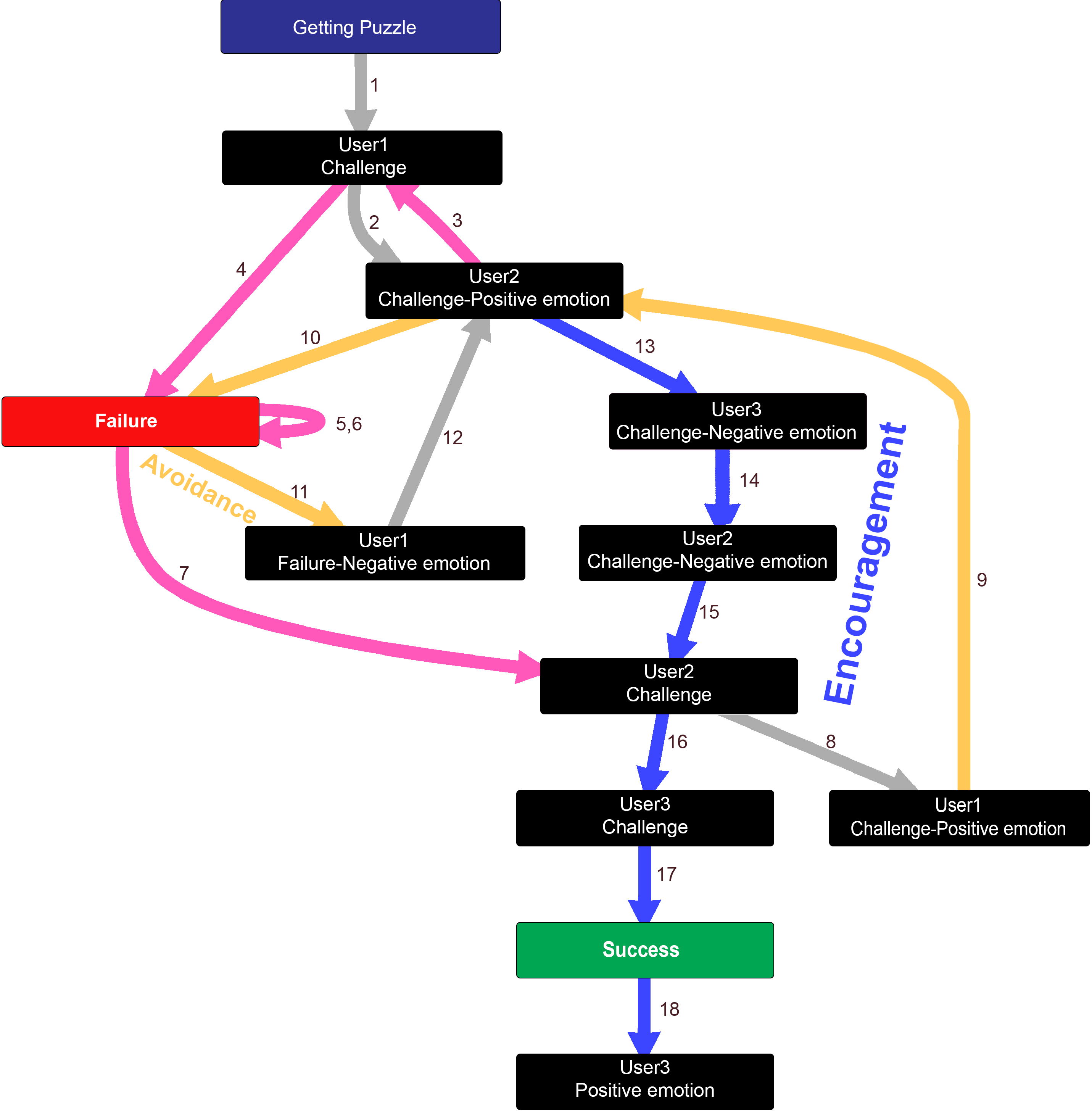}
    \caption{Dependency graph of Team 1 on Puzzle 2 constructed during Phase 3. During Phase 4, emotion regulation events are detected (and highlighted in different colors here).
}
    \label{fig:dependencyGraph}
\end{figure}

\begin{table}[t]
\centering
\begin{tabular}{|p{1.3cm}|p{5.8cm}|}
    \textbf{Labels} & \textbf{Description}  \\
    Failure & Unsuccessful attempts by players to advance toward achieving the game's objectives.  \cite{aytemiz2020diagnostic}  \\
    Conflict & Disagreement regarding an uncertain situation, such as a puzzle solution or a strategy.  \\
    Challenge & Players exposed to a challenging situation, e.g. a newly presented puzzle or an unexpected twist.  \\
    Positive emotion & An emotional reaction designed to express a positive affect, such as happiness.  \\
    Negative emotion & An emotional reaction designed to express a negative affect, such as sadness.  \\
    Getting Puzzle & Receiving a new puzzle.\\
    Success& Correctly solving a puzzle.\\

\end{tabular}
\caption{Events used for Label Classification in Stage 2}
\label{tab:AffectivityFacets}
\end{table}
\subsection{Phase 3: Visualization of Graph Dependency}
Having classified the particular messages into labels, we then 
need to find the most prominent sequences and connections between emotional appearances (e.g., negative emotion, positive emotion) before being able to identify strategies. For this, we draw on sequence graphs, where labels are connected into nodes that represent their respective frequency and edges that connect these nodes in terms of their order. As these are technically closely related to graph representations in process mining, we utilize the process mining package \textit{Disco} for their generation \cite{gunther2012disco}. Figure \ref{fig:dependencyGraph} shows an exemplary dependency graph consisting of nodes (emotion/adaptivity states) and connecting edges (most probable successors of these labels) compiled from the labels of the previous phase. Comparing this to the outlined example in Figure \ref{fig:SixPhases}, User2’s positive message (4:08) lead to the encouragement of User3, which pushed them towards the solution (4:12) and a positive mood eventually (4:13).

In the segment of Figure \ref{fig:dependencyGraph}, the three members of Team 1 employed a multi-tiered approach to problem solving and management regarding puzzle 2. The process map started with “Getting Puzzle” and ended with a positive reaction after solving the puzzle correctly (“Success”). Each node represents a player's sentiment (“positive”, “negative” or “neutral” if not explicitly stated) after events of ``Challenge", ``Failure", and ``Conflict”, connected using directed edges. For example, ``User1 Challenge'' indicates that User1 faced a challenge as puzzle solving and course schedule interfered with each other (“ok I cant go there [right now], not enough time before class, can go there at 5”), while keeping a neutral sentiment. In this particular puzzle, we did not capture any conflict between players.

\subsection{Phase 4: Emotion Regulation Identification}

The generated representations of dependency graphs are then used to identify emotion regulation strategies used by the team in response to failure, challenge or conflict. Using the graphs, we first understand the factors that influence the change of emotional states, where we detected how different situations such as challenge, conflicts, and failure affect players’ interactions. Second, they depict the interplay of different team members with respect to other's emotions. Eventually, we aim to identify paths of player strategies in response to challenge, conflict, and failure that could involve emotion regulation. In order to define player strategies, we utilized Gross' emotion regulation strategies \cite{gross2001emotion}. According to Gross, emotion regulation strategies can be categorized into five families (situation selection, situation modification, attentional deployment, cognitive change and response modulation), which each bears several particular strategies. The emotion regulation cycle usually begins with a difference between a player's goal or desired state and the actual state \cite{gross2015emotion}. This distinction is also known as an ``opportunity for regulation" and it occurs when (1) an alternative strategy is chosen, (2) the strategy is implemented, and (3) the person monitors the success of the regulatory goal.

Example: If we encounter the negative emotion of User3 (4:07) corresponding to an in-game challenge and User2 reacts to this using positive input (4:08), this sequence might contain indications about an emotion regulation strategy that leads to the regulatory goal (in this case, \textit{encouragement}, a strategy of the \textit{situation modification} family, cf. Figure \ref{fig:SixPhases}).

To showcase the identification of emotion regulation patterns, we modified the visualization by assigning a distinct color to each located emotional transition and incremental numbers to the edges (cf. Figure \ref{fig:dependencyGraph}). Using the process model visualization, we identified three major paths as possible indicators of emotion regulation. In the first path (denoted by the color pink), User2 shifted from a positive to a neutral state as a result of attempting to solve the puzzle by guessing the answer and experiencing ``Failure" twice (Path number 3 - 4 - 5 - 6 - 7).
The yellow highlighting denotes the process of User1 working on the solution to the same puzzle afterwards (“User1-Challenge Positive emotion”) by providing hints to User2. However, User2's attempt to provide an answer failed, demotivating User1 so much that they decided to not participate in this puzzle anymore (“User1-Failure Negative emotion”: “wait this is literally chemistry \emoji{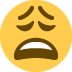}”) after (Path number 9 - 10 - 11), which would correspond to Gross' strategy of \textit{avoidance} \textit{(situation selection)}. Note that the mere execution of such a strategy does not necessarily entail an improvement of emotional states, but we are interested in changes towards both directions.

The third path (blue) represents User3’s impact on User2’s emotional state, as User2 reattempts to construct a solution (“User2-Challenge Positive emotion”: “If this is right then holy shit”) while User3 begins to doubt this reasoning (“User3-Challenge Negative emotion”: “Hmmm. I'm confused“), resulting in a neutral state for both users (“User2-Challenge”: “I'm not sure if what I'm doing is right actually”) (Path number 12 - 13 - 14). Finally, User3 submitted the correct answer after \textit{encouragement}, resulting in an overall transition of a negative to positive sentiment (Path number 14 - 15 - 16 - 17 - 18).

Despite the dependence on manual classification and labeling for this proof of concept evaluation, emotional states, their transitions and the impact of user’s behaviors or game events could already be extracted, captured and visualized. With respect to the remaining teams and puzzles, similar sequences and findings could be identified that lay the ground for subsequent training evaluations of phase 5.

\subsection{Phase 5: Learning Generalizable Concepts}
As soon as multiple instances and segments of similar mechanisms become present, similarities and differences can presumably be learned with sequential models such as recurrent neural networks or comparable approaches that harness internal memory and are capable of extracting sequential commonalities (as depicted in Figure \ref{fig:SixPhases}, Phase 5).
Example: Given (at least) the three segments of the illustrated example, the common strategy of \textit{encouragement} can be learned, which turns out to be the factor that has the highest probability of leading from negative to positive valence in these situations.

\subsection{Phase 6: Situated Intervention}
Building on insights from the prior phases, in-situ occurrences of emotion regulation can be detected within live communication sessions - which enables EAI companions to induce interventions based on already visited or learned reactions in situations where users experience struggle or conflict.
For example, having learned the former labels and mechanisms, a helpful EAI NPC attempts to help overcome the undesirable state of User3 (4:07) by uttering encouraging statements that might bring back User3 towards a positive mood.
In the following, we discuss detailed steps of advancing the formerly manually implemented phases towards automatization and scalability and review possibilities of implementing and evaluating the remaining phases to eventually find generalizable mechanisms of emotion regulation in collaborative conversations and deploy supportive intervention systems into these environments.

\section{Limitations \& Future Work}

There are several limitations to the proposed method. First, the labeling system depends on human labeling which is a laborious process that we aim to address through automation, as discussed below. Another limitation is the sample size, as a considerably larger sample size of data is required to train the intended models around phase 4 and 5. For the purpose of this paper, we restricted the procedure of phase 2 and 4 to a manual proof of concept labeling and classification implementation, which might bias the categorization and will most likely produce different outcomes from established sentiment analysis computation approaches. Among the players in the data set, individual differences (in terms of personality, player type or linguistic manner of expression) should also be taken into account in order to gain a better understanding of each player's emotional state. Another important limitation we aim to extend in future work is the affect of labels. We decided to differentiate only between positive, negative, and neutral sentiment, but aim to investigate deeper psychological dimensions as soon as the scope of collected data and the capability of utilized advanced modeling approaches enables this.

As discussed above, as part of our future work, we strive to automate and connect the particular phases to deliver a fully-fledged tool that turns common raw text data from (collaborative) multi-user conversations into abstractions of affective states and responses that could aid intervention systems and affective, including work that uses CNNs, LSTMs, or RNNs for phase 2. Phase 3 already produces machine-readable sequence graphs but could benefit from fuzzy process mining if the particular segments become too complex for visualization. In this case, robustly configured higher-level parameters for this transformation are vital to ensure compatibility with big variations of size, structure, content and quality between data sets. Phase 4 can already identify sequences between extremes of affective states (i.e. positive to negative transitions and vice versa). In future work, we will explore the use of deeper facets of the human emotion which would necessitate accurate and confident detection mechanisms in order to be usable for finding generalizations in the subsequent step.
According to the results, we identified strategies for emotion regulation by detecting key components in players' behaviors after failures, challenges or conflicts. By collecting these patterns from many players’ gameplay in the future, we are looking forward to train generalizable models in phase 5, such as Attention-Based Bidirectional Long Short-Term Memory (ATTBLSTM) \cite{zhou2016attention}, semi-supervised learning (SSL) approaches like Graph Agreement Models (GAM) \cite{stretcu2019graph} or individual player modeling strategies \cite{pfau2018towards}.
Having an intervention system using the discussed trained model (phase 6) might have several benefits on players such as higher mental health \cite{gross1995emotion}, psychological wellbeing \cite{balzarotti2016individual} as well as players performance and learning. An example of an intervention system in our study could be an artificial intelligence assistant bot that accompanies players throughout a game and, using the trained model, helps players to overcome emotional struggle and spark or amplify positive moods, respectively. This is particularly desirable for the targeted application field of maintaining longer-term motivation in serious game context, such as learning or exercising. Within the context of industrial esports, this endeavor could maintain competitive spirits even after frequent situations of failure, as long it shows plausible behavior that computes efficiently and benefits for the player experience can be quantified \cite{pfau2020case}.
To assess the capabilities and effectiveness of such an intervention system, we are looking forward to evaluate it through usability studies within an educational setting, measuring differences in learning outcomes, usability and player experiences between participants with and without such assistance.
A further example of an intervention system could be a Dynamic Difficulty Adjustment (DDA) system \cite{hunicke2005case}, in which the game is adaptable based on the player's emotional responses \cite{liu2009dynamic}. Finally, since the qualitative data subjects to bias, we will utilize individual differences, and personalized profile for players in order to reduce those biases -- in all stages of data collection, individualized modeling and personalized interaction.

\section{Conclusion}
Emotion regulation denotes a person's affective strategies after stressors or meaningful events and constitutes an important factor for preserving individual resilience. In this paper we presented a data-driven approach to identify emotion regulation strategies that people use when they play team-based games. Our aim is to use this work to develop an AI system that can assist in this coping and emotion regulation process within team-based game settings. 
Identifying such instances of emotion regulation from game chat data is an open problem, which we contribute to with the development of a multi-step approach based on conversational chat log data. We specifically recorded communication data of players playing a team-based game and showcased how the combination of emotion and context identification, dependency graphing of such events, and identification of strategies from these graphs could lead to the detection of emotion regulation. This work is the first step in developing such an assistive EAI. Our next steps are to integrate automated techniques to derive strategies from large datasets. This work blazes the trail for affective support intervention systems towards strengthening resilience in team-based games.
\section{Acknowledgments}
This work is funded by James S McDonnell Foundation (Grant Title: A Methodology for Studying the Dynamics of Resilience of College Students).
\bibliography{aaai22}

\bigskip

\end{document}